\documentclass[conference]{IEEEtran}
\usepackage{cite}
\ifCLASSINFOpdf
\else
  \usepackage[dvips]{graphicx}
  \graphicspath{{../eps/}}
  \DeclareGraphicsExtensions{.eps}
\fi
\usepackage{fixltx2e}
\usepackage{stfloats}
\usepackage{url}
\usepackage[T1]{fontenc}
\usepackage{listings}
\lstdefinelanguage{p4}
{ morekeywords={*,extern_type, attribute, type, method, extern, action, control, void},
  sensitive=true,
  morecomment=[l]{//}, % l is for line comment
  morecomment=[s]{/*}{*/}, % s is for start and end delimiter
  morestring=[b]" % defines that strings are enclosed in double quotes
}

\usepackage{float}
\usepackage{enumitem}

\usepackage{balance}
\usepackage{subcaption}
\usepackage{pgfplots}
\usepackage{pgf}
\usepackage{tikz}
\usetikzlibrary{patterns}
\begin{document}
\title{Extern Objects in P4: an ROHC Header Compression Scheme Case Study}

\author{\IEEEauthorblockN{Jeferson Santiago da Silva, Fran\c{c}ois-Raymond Boyer, Laurent-Olivier Chiquette and J.M. Pierre Langlois}
\IEEEauthorblockA{Department of Computer and Software Engineering\\
Polytechnique Montr\'{e}al, Canada\\
Email: \{jeferson.silva, francois-r.boyer, laurent-olivier.chiquette, pierre.langlois\}@polymtl.ca}
%\and
%\IEEEauthorblockN{Homer Simpson}
%\IEEEauthorblockA{Twentieth Century Fox\\
%\and
%\IEEEauthorblockN{James Kirk\\ and Montgomery Scott}
%\IEEEauthorblockA{Starfleet Academy\\
%San Francisco, California 96678--2391\\
%Fax: (888) 555--1212}
}


%Telephone: (800) 555--1212, Fax: (888) 555--1212}


\maketitle

\begin{abstract}
P4 is an emergent packet-processing language with which the user can describe how the packets are to be processed in a switching element. This paper presents a way to implement complex operations that are not natively supported in P4. In this work, we explored two different methods to add extensions to P4: i) using new native primitives and ii) using extern instances. As a case study, an ROHC entity was implemented and invoked in a P4 program. The tests showed similar relative performance in both methods in terms of normalized packet latency. However, extern instances appear to be more suitable for target-specific switching applications, where the manufacturer/vendor can specify its own specific operations without changes in the P4 syntax and semantics. Extern instances only require changes in the target-specific backend compiler while keeping the P4 frontend compiler unchanged. The use of externs also results in a more elegant code solution since they are implemented outside the switch-core, thus reducing side effects risks that can be caused by a modification in a switch pipeline implementation. 
\end{abstract}


% \fi
\IEEEpeerreviewmaketitle

\section{Introduction}\label{sec:intro}
%% no \IEEEPARstart
%This demo file is intended to serve as a ``starter file''
%for IEEE conference papers produced under \LaTeX\ using
%IEEEtran.cls version 1.8b and later.
%% You must have at least 2 lines in the paragraph with the drop letter
%% (should never be an issue)
%I wish you the best of success.

SDN promises to fill the gaps of adaptability and scalability of current rigid network implementations. SDN networks decouple the data- and the control-planes, abstracting the low-level hardware implementations. Thus, in SDN-based networks, one major challenge is how to represent the data-plane implementation in a programmable and portable fashion using off-the-shelf commodity forwarding elements (FEs), known to be hard to program. Several high-level packet processing languages have been developed in recent years \cite{DUNCAN:09,Foster:11,Song:13,Foster:13,Bonelli:14} to address this issue.

In this work, we have explored the capabilities of P4, a protocol- and target-independent packet processing language. P4 describes how the packets are to be processed in an FE, such as a switch or a router. With P4, it is easily possible to implement the de facto SDN protocol: OpenFlow \cite{theopennetworkingfoundation2014}. This is because a P4 program specifies at a high-level which set of headers are processed in an FE. This adaptability characteristic makes P4 suitable for SDN-based networks, permitting new protocols to be easily deployed in an FE without changing the underlying hardware device. 

The main goal of this work is to propose new extensions in P4. In our work, we have explored two different ways to add new P4 commands: i) using new native P4 primitives, and ii) using extern method calls, which required modifications to the P4 frontend and backend compiler as well as to the P4 switch model. The proposed modifications to the backend compiler are now publicly available in the P4 repository \cite{p4_git:16}. We also propose a new native primitive to P4: \textit{modify\_and\_resubmit} a packet.

We used a RObust Header Compression (ROHC) scheme as a P4 extensions case study. The tasks required by the ROHC entity have driven the implementation of new extensions to P4, since only using the current language constructs would be very hard, if not impossible, to describe the entire header compression/decompression process.

In this work, we tried to reuse as much as possible the P4 infrastructure, trading-off design re-usability, and cost of coding. The additional constructs were implemented in C/C++ and they were called directly in the P4 program as new P4 constructs. The base for this work are the P4 compiler and behavioral model switch implementation \cite{p4_git:16} and an open-source Linux ROHC library \cite{Barvaux:16}. Our implementation is also available in open-source on GitHub\footnote{\url{https://github.com/engjefersonsantiago/p4-programs}}.

The rest of this paper is organized as follows: Section~\ref{sec:related_works} presents a review of the literature,  Section~\ref{sec:method} draws the methodology adopted in this work, Section~\ref{sec:p4_rohc} presents the required P4 support for the ROHC scheme, Section~\ref{sec:results} shows the experimental results and discussions, and, Section~\ref{sec:concl} draws the conclusions and future works. 

\section{Related Works}\label{sec:related_works}

SDN-based networks have emerged in recent years as a viable solution to deploy switched networks worldwide. Since in SDN the control and data planes are decoupled, FE appliances are simplified, performing only a set of fixed actions applied to the packet headers based on pre-defined matching rules, configured by a centralized controller. 

Packet processing languages leverage the SDN paradigm. They ease the network programmability by using a network-specific programming dialect, rather than general-purpose programming languages. Recent efforts have tackled domain specific languages for network applications, with significant adoption of two packet processing languages: POF, and more recently, P4. In \cite{Song:13}, Song present the POF language: a protocol-oblivious packet processing language targeting network processors. As an evolution of his previous work, Song \textit{et al.} \cite{song2015unified} introduce the concept of abstract forwarding model to expand POF support to a variety of hardware architectures.

\begin{figure}[H]
\centering
\includegraphics[width=3.3in]{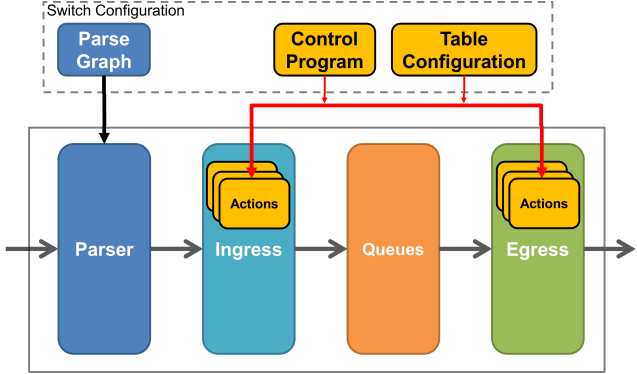}
% where an .eps filename suffix will be assumed under latex, 
% and a .pdf suffix will be assumed for pdflatex; or what has been declared
% via \DeclareGraphicsExtensions.
\caption{P4 Abstract Forwarding Model.}
\label{fig:p4_abs_model}
\end{figure}

In \cite{Bosshart:14}, Bosshart \textit{et al.} propose the P4 language, a protocol independent packet processing language. Fig.~\ref{fig:p4_abs_model} presents the abstract forwarding model on which P4 was built upon. A P4 program consists of a parser state machine (PSM) followed by a set of match-action tables in the ingress and egress pipelines. The processing flow is controlled by an imperative control program.

%\begin{figure}[H]

The state transitions in the PSM are driven by the value of header fields. In P4, the headers are defined in terms of their fields in a \textit{header} structure (similar to a \textit{struct} in C), thus avoiding error-prone bit-level manipulation. For the match-action tables, four matches types are supported: exact, range, ternary and wildcard. P4 actions execute as procedures composed of native P4 primitives. The control program defines in which order tables are applied according to imperative statements.

Due to its simplicity, portability, and device agnosticism, P4 has gained popularity as a packet processing language for programmable forwarding elements in both academia and industry \cite{Abdi:2016,Li:2016,Shahbaz:2016,Pfaff:2015,Wang:2017}.

Some works in the literature have proposed to add extensions to P4. Sivaraman \textit{et al.} \cite{Sivaraman:2015} propose a case study for the utilization of P4 switches for data center applications. The authors map P4 deficiencies and propose new primitive actions to P4. However, the new primitives are placed within the P4 switch core, not using external libraries. Huynh \textit{et al.} \cite{Huynh:2016} have reported the use of target-specific externs with a custom version of P4. They used a proprietary backend compiler \cite{Netronome:2017} toolset to generate low-level custom code for NICs from the P4 description. 

In this context, it is clear that the integration of P4 and extern standard libraries is a potentially powerful way to describe target-specific packet processing operations at the user abstraction level. Moreover, there is a lack of available non-commercial backend compilers to support externs described using the standard P4 syntax and semantics.

\section{Methodology}\label{sec:method}

This section presents the methodology applied in this work. In P4 it is possible to add new primitives in two different ways: as a new native primitive or through extern instances. The two approaches are presented in the following sections.

\subsection{P4 Extensions: New Native Primitives}\label{sec:p4_nat}
The P4 consortium provides a behavioral switch implementation of its abstract switch model described in C++. The behavioral model is configured through a JavaScript Object Notation (JSON) file generated by the P4 backend compiler. The P4 backend is fed by a collection of data structures representing the P4 intermediate representation (IR) generated by the P4 frontend compiler. The set of native primitives is defined in an input JSON array, that feeds the P4 frontend compiler. This JSON array has a corresponding description for the primitive actions in the P4 switch model, including function names and parameters.

To add new primitives to P4, the user must edit the primitives JSON description, by adding the new primitive and its parameters. The behavioral model, in turn, must support the new primitive. Each P4 switch target can have its own set of primitives. In the behavior model, these primitives are defined in a file named \textit{primitives.cpp}\footnote{https://github.com/engjefersonsantiago/behavioral-model}. This code implements the behavior of all supported P4 primitives, described as C++ functors. To add the new primitive, a new functor must be included in this file, with the same name as the new P4 primitive followed by the macro \textit{BM\_REGISTER\_PRIMITIVE(my\_p4\_primitive)}.

\subsection{P4 Extensions: Extern Instances}\label{sec:p4_ext}

P4 already supports extern methods in the switch model and in the P4 frontend compiler in the most recent P4-16 version \cite{p4:16}. However, the JSON file generation was not implemented and therefore not integrated with the P4 backend compiler. The P4 version 1.1.0 \cite{p4v1.1:14} had also implemented a draft for extern objects, but this version was deprecated. Since P4-16 only came to the public recently, much of the present work was done in the version 1.1.0 and afterwards ported to P4-16. A 90 lines-of-code (LOCs) patch in the P4 v1.1.0 backend compiler\footnote{https://github.com/engjefersonsantiago/p4c-bm} was enough to generate the correct JSON arrays based on the P4 extern description. The equivalent support in the P4-16 backend\footnote{https://github.com/engjefersonsantiago/p4c} required as few as 50 LOCs. Listing~\ref{list:p416_desc} shows the P4 description of an extern in P4-16.

%\scriptsize
\begin{lstlisting}[language=p4,basicstyle=\footnotesize,caption=Description of a P4 Extern Instance., label=list:p416_desc,frame=lines,basicstyle=\footnotesize\ttfamily]
extern extern_example {
  ext_type(bit<1> attribute_example); 
  void method_example();
}

control control_example {
  extern_example(0x0) my_extern_example;

  action my_extern_call(){
    my_extern_example.method_example();
  }
}
\end{lstlisting}

In Listing~\ref{list:p416_desc}, the P4 reserved word \textit{extern} defines the type of the extern instance. Object constructors are used for passing the initialization parameters to the extern type. To finalize the extern type definition, the list of the supported extern functions is defined, and each method can include several parameters, such as metadata, header fields, and constant integer values. The extern instance is declared as objects in C++, including the initialization parameters. The extern methods are accessed by a given instance using an object-oriented notation: \textit{extern\_instance\_name.method\_a(method\_parameters ...)}, which is called inside a P4 action. Listing~\ref{list:json_desc} represents the equivalent JSON array of the declared extern P4 object.

\begin{lstlisting}[float,basicstyle=\footnotesize,label=list:json_desc,caption=JSON Array of an Extern Instance., frame=lines,basicstyle=\footnotesize\ttfamily]
"actions": [{
    "name": "my_extern_call",
    "id": 0,
    "runtime_data": [],
    "primitives": [{
        "op": "_extern_example_method_example",
        "parameters": [{
            "type": "extern",
            "value": "my_extern_example"}]}]}]
"extern_instances": [{
    "name": "my_extern_example",
    "id": 0,
    "type": "extern_example",
    "attribute_values": [{
        "name": "attribute_example",
        "type": "hexstr",
        "value": "0x0"}]}]
\end{lstlisting}

For proper operation of the extern methods in P4, the target architecture must include the extern modules. As the P4 behavioral model is a soft-switch, the extern types can be described in C++ and instantiated in the switch core. The extern module uses the P4 behavioral model action class to register the extern methods as new P4 primitives in the switch. Therefore, when these extern actions are reached in a given switch pipeline, the P4 model knows this is an external call, and invokes the correct extern method. The Listing~\ref{list:cpp_extern} shows the equivalent switch implementation for the extern type declared above in P4.

\begin{lstlisting}[language=C++,morekeywords={*,override},float,basicstyle=\footnotesize,label=list:cpp_extern,caption=C++ Class of an Extern Instance., frame=lines,basicstyle=\footnotesize\ttfamily]
#include <bm/bm_sim/extern.h>
using namespace std;
template <typename... Args>
using ActionPrimitive =
   bm::ActionPrimitive<Args...>;
using bm::Data;
class extern_example : public ExternType {
 public:
  BM_EXTERN_ATTRIBUTES {
    BM_EXTERN_ATTRIBUTE_ADD(attribute_example);
  }
  void init() override {}
  void method_example () {
    cout << "Dummy extern method call\n";
  }
 private:
  Data attribute_example;
};
BM_REGISTER_EXTERN(extern_example);
BM_REGISTER_EXTERN_METHOD(extern_example,
                          method_example);
int import_extern_example(){return 0;}
\end{lstlisting}

The extern class is declared as a child class of the \textit{ExternType}, provided by the P4 behavioral model, and it is compiled along with the target switch. The macros \textit{BM\_EXTERN\_ATTRIBUTES} and \textit{BM\_EXTERN\_ATTRIBUTE\_ADD} are used to link the C++ initialization variables with those that are defined in the P4 instantiation and passed to the switch model as a JSON array. The \textit{init} function is a non-accessible P4 method, it is mandatory and it is used to initialize the stateful classes and variables of the extern instance. The list of methods follows the \textit{init} function definition. The extern type is registered in the switch model through the macro \textit{BM\_REGISTER\_EXTERN} and the methods are set as new P4 primitives by the macro \textit{BM\_REGISTER\_EXTERN\_METHOD}. A dummy function is then declared and it must be called in the target, as extern, to force linking of that translation unit.

\section{ROHC Support in P4}\label{sec:p4_rohc}

Drafts for next-generation cellular communication, such as 5G, assume that mobile network equipment will exploit concepts of SDN and NFV, including base-stations and forwarding elements \cite{Bradai:15,Ramirez:16}. SDN-aware devices are expected to be present in the wireless backbone, probably supporting some packet processing language, such as P4. Due to this, we chose to implement a header compressing scheme in a P4 switch as a case study of extern objects. This choice was based on the fact that header compression is a requirement for LTE systems and it appears as a requirement for the forthcoming 5G. Many header compression schemes have been standardized in the last 20 years. In this work, we focus on the ROHC scheme \cite{rfc4995}, the standard for LTE systems. The next two subsections respectively present the requirement analysis for an ROHC entity and the final implementation architecture we decided to carry out in this work.

\subsection{ROHC Implementation Requirements}

Header compression schemes reduce the overhead caused by large headers by not sending with each packet redundant or easily predictable header information. Considering a flow of packets belonging to the same application, many header fields can be considered static, such as addresses and ports, and these fields do not need to be transmitted at all. Other fields change in a predictable way, such as sequence numbers, allowing the compressor entity to only transmit the difference between the current and the previous packet belonging to the same flow. Fields such as checksums and CRCs cannot be compressed and should be transmitted as is.% Fig.~\ref{fig:rohc_scheme} illustrates a data transfer using a header compressing scheme over an Ethernet network.

%%\begin{figure}[!t]
%\begin{figure}[!htb]
%\centering
%\includegraphics[width=3.3in]{./figures/rohc_scheme_n}
%
%% where an .eps filename suffix will be assumed under latex, 
%% and a .pdf suffix will be assumed for pdflatex; or what has been declared
%% via \DeclareGraphicsExtensions.
%\caption{ROHC compressor scheme.}
%\label{fig:rohc_scheme}
%\end{figure}

Packet header compression/decompression involves several steps. Among them, we highlight packet profiling and context maintenance. Profiling is related to which type of compression technique will be applied to the headers, depending on the protocols involved in the communication. The ROHC standard defines up to 15 profiles supported for header compression. Profiling packets are easily done in P4 through the PSM, where a profile identifier is assigned according to the received protocols.

Context maintenance is a more complex task, where new packet flows must be assigned to a new data context, while similar packet flows are analyzed in order to determine whether they belong to an existent context. This context manipulation involves the matching operation of several header fields, depending on the profile. P4 allows the user to perform several match types in tables, including exact match. However, P4 does not support table elements insertion in the data-plane, which is necessary for context creation. 

As an option, context tables can be implemented using the available P4 register construction. However, registers do not support matching operations. Implementing such a match table behavior in registers is a complicated task. One solution would be to use a hash function to index an $N \times$1-bit register. This $N \times$1-bit register plays the role of a match bit for a table of $N$ entries. The same hash value would be used to index other registers holding the context. A basic limitation of this approach is the high risk of conflicts due to the bit limited key (up to 14 bits) used to encode several match fields. In addition, the decompressor needs to maintain several recent contexts belonging to the same flow identifier, in order to recover the packet in case of a recent packet drop. Due to these difficulties, in this work the context tables and matching engines were kept outside the P4 implementation, copying the current packet to the ROHC compressor and decompressor entities.

ROHC applies different header compression encoding techniques over the headers, including LSB and slide window LSB encoding, and scaled timestamp encoding. These encoding schemes along the multiple ROHC packet formats result in a variable packet size, which is determined when the compression operation ends. P4 does a good job with fixed sized header formats, the ones the P4 PSM must know in advance in order to extract the needed fields or for being part of the parser graph for further correct packet deparsing. Thus, using native P4 constructs, the simple task of parsing an ROHC header is not as easy as it may seem. An alternative solution we used is to interpret the ROHC header as part of the packet payload, which is done entirely in the P4 behavioral model.

\subsection{Adopted ROHC Implementation}

\begin{figure}[!htb]

\centering
\includegraphics[width=3.3in]{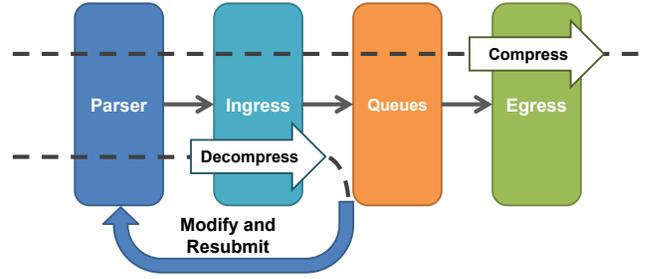}
% where an .eps filename suffix will be assumed under latex, 
% and a .pdf suffix will be assumed for pdflatex; or what has been declared
% via \DeclareGraphicsExtensions.
\caption{Modified P4 Abstract Forwarding Model: ROHC Support.}
\label{fig:p4_modified_pipeline}
\end{figure}

To support the ROHC compression and decompression schemes, we statically compiled an ROHC library and instantiated it in the modified behavior switch model. Performing header compression/decompression required three new primitives in P4: \textbf{ROHC compression}, \textbf{ROHC decompression}, and \textbf{modify and resubmit} a packet. Fig.~\ref{fig:p4_modified_pipeline} shows the proposed modified switch model.

%\begin{figure}[H]

The compression operation starts in the PSM when a packet arrives at the switch. The PSM, described in P4, sweeps all headers in order to identify which ROHC profile is going to be compressed, storing it into packet metadata. This metadata is then used by the new P4 primitive, represented in P4 as \textit{rohc\_comp\_header}. The extracted uncompressed headers are translated into an acceptable ROHC data structure and passed to the ROHC compressor entity. The compressor engine performs the operation, returning a compressed byte stream, that is placed at the beginning of the packet payload. The original packet headers are invalidated and they are not transmitted. The last step in the compression task is to differentiate the ROHC packets at the lowest layer protocol. As this work is based on Ethernet networks, an unused Ethertype value was assigned for ROHC packets for proper packet identification.

Similar to the compression, the decompression is triggered in the PSM, where the parser identifies the special Ethertype. Once identified, the decompression takes place. A very similar data structure conversion is applied to the packet payload before passing it to the decompressor engine, named \textit{rohc\_decomp\_header} in P4. The decompressor then performs full headers recovery, placing them at the beginning of the payload field. The last steps in the decompressor are to remove the former compressed header and to assign a valid Ethertype value in the Ethernet header: 0x0800 (IPv4).

Since P4 describes how a packet is processed and forwarded, the operations of header compression and decompression have no value if the forwarding rules cannot be applied to the packets. Let us consider a decompressing operation: i) a packet arrives at the switch; ii) the packet is identified as an ROHC packet; iii) let us assume the packet is correctly decompressed; iv) the decompressed packet is processed and forwarded according to the packet rules; and v) the packet is transmitted. However, for step iv), the packet headers must be part of the parser graph in order to be correctly extracted by the switch. To handle this, it is necessary to re-parse the recovered uncompressed packet. The solution adopted in this case was to send it back to ingress parser. For this, a new P4 primitive was created, \textit{modify\_and\_resubmit}, instead of using the existing \textit{recirculate} primitive, aiming to reduce the packet latency, since the packet recirculation primitive is only applied at the end of the egress pipeline.

For the \textit{modify\_and\_resubmit} primitive, the ingress pipeline was modified in the P4 switch model. The existing \textit{resubmit} primitive forwards back a packet to the parser. However, all the modifications performed on the packet in the ingress pipeline are not applied and a clone of the original packet is sent back to the parser. To allow the packet modifications, we added a deparser entity before resubmitting the packet, that takes the updated header fields and the packet payload, and serialize them into a stream of bytes. For proper packet resubmission, the packet length was also updated and stored in the switch standard metadata. 

\section{Experimental Results and Discussions}\label{sec:results}

\begin{figure}[!htb]
\centering
\includegraphics[width=3.3in]{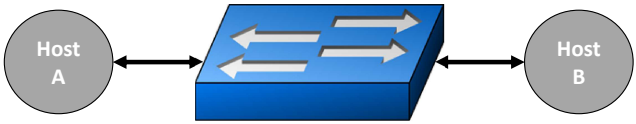}
% where an .eps filename suffix will be assumed under latex, 
% and a .pdf suffix will be assumed for pdflatex; or what has been declared
% via \DeclareGraphicsExtensions.
\caption{Emulated Network Topology.}
\label{fig:topo}
\end{figure}

To validate our modifications in the P4 compiler and in the behavioral model switch, we created an emulated three-node network. Fig.~\ref{fig:topo} illustrates the network topology. This simple network is composed of two hosts connected to each other through a P4 switch. A Python script was used to send packets from host A to host B, passing through the switch. The packets sent by host A are compressed. The switch then uncompresses the packets, performs the forwarding procedure, recompresses the packets and forwards the packets to B.

%\begin{figure}[!ht]

%The new P4 primitives were tested in four different scenarios: i) as a native primitive included directed in the P4 behavioral model, ii) as extern methods being instantiated in the  P4 soft-switch. The scenarios iii) and iv) are particular cases of both solutions i) and ii), where the switch is configured to recirculate packets, instead of using the new modify and resubmit primitive. During the tests, all the solutions showed to be equivalent in terms of functionality. In the case of the solutions i) and iii), an intrinsic drawback is that any modification in the new primitive requires a recompilation of the whole set of primitives since they belong to the same C++ file in the switch model.

The new P4 primitives were tested through the following scenarios:
\begin{itemize}[noitemsep,topsep=0pt]
\item using new native primitives;
\item using extern methods.
\end{itemize}

The native primitives were included directly in the P4 behavioral model, while the extern methods were instantiated in the P4 soft-switch. Both scenarios were tested in two different conditions. One using the current recirculate primitive of P4 to send the packets back to the PSM, and the other using the new modify and resubmit primitive. It is important to highlight that both recirculate and modify and resubmit primitives are equivalent in terms of functionality, the difference between them resides in which pipeline stage the action is applied.

The tests carried out in this section are used to assure the viability of our method, presenting a comparison trade-off between the methods we developed. Therefore, in this section, we used as a comparison metric the normalized average packet latency. The baseline for this metric is the ROHC implementation using new native P4 primitives with the original P4 packet recirculate primitive, because this implementation is the most similar to the original P4 soft-switch.

\begin{figure}[!htb]
\centering
\begin{tikzpicture}
\begin{axis}[
    legend columns=2,
    legend entries={\tiny Primitive + Recirculate, \tiny Primitive + Modify and resubmit, \tiny Extern + Recirculate, \tiny Extern + Modify and resubmit},
    legend to name=CombinedLegendBar,
    footnotesize,
    area legend,
    xtick=data,
    xticklabels,
    %nodes near coords,
    ymin=0,
    ymax=1.05,
    bar width=0.05\textwidth,
    x=1.5cm, % Distance between the centers of the bars
    enlarge x limits={abs=1cm}, % The distance between the center of the first bar and the left edge
    enlarge y limits=false,
    height=0.33\textwidth,
    width=0.5\textwidth,
  ]
    \addplot[ybar, pattern=vertical lines] coordinates {(1, 1.00)};
    \addplot[ybar, pattern=north east lines]   coordinates {(2, 0.79)};
    \addplot[ybar, pattern=grid]             coordinates {(3, 1.01)};
    \addplot[ybar, pattern=dots]             coordinates {(4, 0.82)};
\end{axis}
\end{tikzpicture}
\ref{CombinedLegendBar}
\caption{Normalized Packet Latency.}
\label{fig:pkt_latency}
\end{figure}
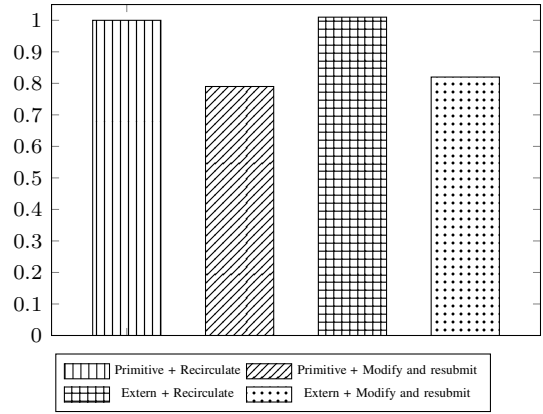

For testing purposes, we generate a set of 10,000 compressed ROHC packets, more than sufficient to show statistically significant differences in normalized execution times between our scenarios. The original uncompressed packets are 74 bytes length, including 20 bytes of RTP payload. The normalized packet latency presents variations according to the test scenario and condition. These results are summarized in Fig.~\ref{fig:pkt_latency}. The considered packet latency metric is the time taken to receive an uncompressed packet, perform the packet modifications, recompress and forward it to the destination host.

%%\begin{table}[!ht]
%\begin{table}[!htb]
%% increase table row spacing, adjust to taste
%\caption{Normalized Packet Latency.}
%\centering
%% Some packages, such as MDW tools, offer better commands for making tables
%% than the plain LaTeX2e tabular which is used here.
%\begin{tabular}{ccc}
%\hline
%\textbf{Extern Object}  & \textbf{Modify and Resubmit} & \textbf{NPL}\\
%\hline
%                 &                              & 1.00\\
%                 & $\surd$                      & 0.79\\
%$\surd$          &                              & 1.01\\
%$\surd$          & $\surd$                      & 0.82\\
%\hline
%\label{table:pkt_latency}
%\end{tabular}
%\end{table}

%\begin{figure}[!htb]
%\caption{Normalized Packet Latency.}
%\begin{tikzpicture}
%\begin{axis}[
%    symbolic x coords={Primitive + Recirculate, Primitive + Modify and resubmit, Extern + Recirculate, Extern + Modify and resubmit},
%    xtick=data,
%    ymin=0,
%    ymax=1.05,
%    bar width=0.05\textwidth,
%    width=0.5\textwidth,
%    ylabel=Normalized Packet Latency,
%  ]
%    \addplot[ybar,fill=blue] coordinates {
%        (Primitive + Recirculate, 1.00)
%        (Primitive + Modify and resubmit, 0.79)
%        (Extern + Recirculate, 1.01)
%        (Extern + Modify and resubmit, 0.82)
%    };
%    
%\end{axis}
%\end{tikzpicture}
%\label{fig:pkt_latency}
%\end{figure}

Considering the data from Fig.~\ref{fig:pkt_latency}, the normalized latency is lower by 21\% when using the native primitive with the modify and resubmit primitive. Under the same condition but using externs, the latency is reduced by 18\%. The egress pipeline processing latency can explain the latency reduction in the scenarios that utilize the new modify and resubmit primitive. This new primitive action is applied in the ingress pipeline and then the packet does not traverse the egress pipeline before being sent back to the parser. 

Even though native primitive scenarios slightly outperform extern objects, this approach introduces a compatibility and modularity problem. Proposing extensions through this method is not feasible in the standard P4 language because each new primitive requires frontend compiler support (and target-specific backend). The use of extern objects presents a modular and scalable way to describe target-specific features with P4 using target-specific P4 externs defined in a standard library. Externs instances are an elegant yet portable solution since the P4 frontend compiler remains unchanged regardless of whether externs are used or not. Implementing externs is a backend compiler task, the only changed toolset in the design tool.

In terms of code complexity, code style, and design re-usability, externs offer with no doubt a much more elegant solution. In the P4 code, the externs are seen as target-specific objects, keeping then the natural simplicity of the language. When extending the language with new native methods, the user must pay attention to all already implemented primitives, providing support to the new ones in the P4 frontend compiler, which is undoubtedly a difficult and arduous task. From the P4-switch viewpoint, using extern objects is as simple as instantiating a new class in an object-oriented program. With native primitives, the new actions must be developed inside the switch-core modifying the original switch pipeline design, which can lead to user applications misbehavior.

It is also important to highlight that adding new native primitives to P4 incurs increasing switch complexity. All native primitives belonging to the standard P4 language are supposed to be supported by all P4-compatible switch targets. Since the deprecated P4 v1.1.0 version \cite{p4v1.1:14}, the P4 consortium has attempted to reduce the number of native primitives aiming to keep the language as concise as possible, only implementing basic operations. It was standardized in P4-16 \cite{p4:16}, where most constructs that were once part of P4 have been redefined as externs. 

\section{Conclusion and Future Work}\label{sec:concl}
In this work, we presented how to add new extensions to P4. Two different methods were presented: by adding new native primitives and using extern methods. Both methods were tested using the simple switch target provided by the P4 Consortium and they showed to be effective. As a case study, an ROHC compressing entity was used due to the relevance of compression techniques in the field of mobile cellular communications, a field that is expected to grow in SDN adoption in next years. Both compressor and decompressor engines were integrated into a P4 program, where the user can decide whether or not to use the header compression scheme. 

We highlight in this work the use of extern instances in P4 and the possibility of its integration with proprietary libraries, with the objective of keeping the natural simplicity of the language. Through extern methods, the switch manufacturers can develop their more advanced operations the way it is more convenient for their applications, such as a specialized software implementation or a hardware accelerator, while requiring no modifications in the P4 frontend compiler.

As future work, we aim to explore other P4 compatible switch platforms and expand the use of extern instances on these platforms. It is also an objective of future research to use hardware accelerators described as extern methods in P4 in order to perform timing critical operations, such as packet classification, traffic management, and deep packet inspection.

\section*{Acknowledgments}

We would like to thank the P4 consortium members for their technical support. This work was carried out with the support of the Brazilian National Council for Scientific and Technological Development (CNPq), Natural Sciences and Engineering Research Council of Canada, and Ericsson Research Canada.

\balance



\bibliographystyle{./IEEEtranBST2/IEEEtran}
\bibliography{./IEEEtranBST2/IEEEabrv,./IEEEtranBST2/IEEEexample}

%\begin{thebibliography}{1}
%\end{thebibliography}

\end{document}